\documentclass[showpacs,aps,preprint, prd,nofootinbib,floatfix,amsmath,amssymb]{revtex4-1}
\usepackage{graphicx}  % needed for figures
\usepackage{amsmath,amssymb,bm}
\usepackage{amsmath}
\usepackage{xcolor}
\def\beq{\begin{eqnarray}}
\def\eeq{\end{eqnarray}}

\begin{document}

 \title{Flavor Physics constrains on a $\mathbb{Z}_5$-3HDM}

\author{Alfredo Aranda$^{1,2}$\footnote{fefo@ucol.mx}, J. E. Barradas-Guevara$^{3}$\footnote{barradas@fcfm.buap.mx}, A. Cordero-Cid$^{4}$\footnote{luzadriana\_cordero@hotmail.com}, Francisco de Anda$^{5}$\footnote{franciscojosedea@gmail.com}, Antonio Delgado$^{6}$\footnote{antonio.delgado@nd.edu}, O. F\'elix-Beltr\'an$^{4}$, J. Hern\'andez-S\'anchez$^{2,4}$\footnote{jaime.hernandez@correo.buap.mx}} 
\affiliation{$^1$Facultad de Ciencias - CUICBAS, Universidad de Colima, Colima, M\'exico,\\
$^2$Dual $C$-$P$ Institute of High Energy Physics, M\'exico. \\
$^3$Facultad de Ciencias F\'{\i}sico Matem\'aticas, Benem\'erita Universidad Aut\'onoma de Puebla, Apdo. Postal 1152, Puebla, Puebla, M\'exico.\\
$^{4}$Facultad de Ciencias de la Electr\'onica, Benem\'erita Universidad Aut\'onoma de Puebla, Apdo. Postal 542, C.P. 72570, Puebla, Puebla, M\'exico. \\
$^{5}$Departamento de F\'{\i}sica, CUCEI, Universidad de Guadalajara, M\'exico.\\
$^{6}$Department of Physics, University of Notre Dame, Notre Dame IN 46556, USA.}

\date{\today}

\begin{abstract}
We present the general behavior of the scalar sector in a Three Higgs Doublet Model (3HDM) with a $\mathbb{Z}_5$ flavor symmetry.  There are regions of the parameters space where it is possible to get a SM-like Higgs boson with the other Higgs bosons being heavier, thus decoupled from  the SM, and without relevant contributions to any flavor observables. There are however other more interesting regions of parameter space with a light charged Higgs ($m_{H^\pm} \sim $ 150 GeV) that are consistent with experimental results and whose phenomenological consequences could be interesting.  We present a numerical analysis of the main $B$-physics constraints and show that the model can correctly describe the current experimental data.
\end{abstract}


\maketitle
\section{ \label{sec:introduction} Introduction}
The discovery of the Higgs at CERN has opened a new era in particle physics. For the first time in history we are sure that the breaking of the electroweak symmetry is triggered by the vacuum expectation value of a scalar field. 

One thing that is still unanswered is whether there is just one or several Higgs fields. In this paper we go through the second avenue and 
study the flavor effects present in the Three Higgs Doublet Model (3HDM) with cyclic flavor group $\mathbb{Z}_5$ given in Ref. \cite{Aranda:2013kq}.
This particular flavor symmetry is well  motivated since it is the smallest Abelian symmetry that enables the Nearest Neighbor Interaction (NNI) Yukawa textures in the quark sector of the model.  Furthermore, the lepton sector of the model  is completely determined by the flavor symmetry, with no right-handed neutrinos, and Majorana neutrino  masses being generated radiatively through the presence of a SU(2) singlet field charged under both hypercharge and lepton number.

An important issue that any model with an extended Higgs sector must address is the existence of strong experimental constraints that both the Higgs data and, in this case, the flavor data can impose in its parameter space. The goal of this paper is to present a detailed analysis of both the Higgs and flavor sectors of the above-mentioned model and see if there are regions of the parameter space that can lead to a light scalar spectrum testable at the LHC.

The paper is organized as follows: the Higgs potential is studied in section~\ref{sec:higgspotential} with special emphasis to the constraints coming from $K-\overline{K}$ mixing; the Yukawa sector is then analyzed in section~\ref{sec:yukawasector} imposing the constraints coming from $h\to\gamma\gamma$ and then in section~\ref{sec:flavorconstraints} the core results of the paper are presented showing the different contributions to the different exotic decays. Finally section~\ref{sec:conclusions} is devoted to our conclusions.

\section{Higgs potential}\label{sec:higgspotential}

In this model the Higgs sector consists of three SU(2) doublet fields denoted by ${\cal{H}}=(H,\Phi_1,\Phi_2)$ and a singlet scalar field $\eta$ with $Y=\frac{1}{2},-1$ respectively. 
The matter content is as follows: left-handed doublets fermionic fields ${Q}_L,{L}_L$ and right-handed singlet fermionic fields $u_R, d_R$ and $e_R$. Each sector  has a $\mathbb{Z}_5$-charge assignment for each family: for the fermionic fields $\bar{Q}_L \simeq \bar{L}_L \simeq (0,-3,-1)$, $u_R \simeq d_R \simeq e_R \simeq (3,0,2)$; and for the scalar sector ${\cal{H}}(H,\Phi_1,\Phi_2) \simeq (0,-1,1)$, $\eta \simeq(-1)$.  These assignments lead to NNI textures for the quarks and charged lepton Yukawa matrices, and two zeroes of texture for the neutrino mass matrix. 

The Higgs potential is given by \cite{Aranda:2012bv}
\begin{equation}
\begin{array}{rcl}
V(H,\phi_a)&=&\mu^2_0 {\mid H \mid}^2+\mu^2_a {\mid \phi_a \mid}^2+\mu^2_{0a} ({\phi_a}^{\dagger} H+{\it{h.c.}})+ \lambda_0( {\mid H \mid}^2)^2 +\lambda_a( {\mid \phi_a \mid}^2)^2\cr
&&+\lambda_{0a} {\mid H \mid}^2 {\mid \phi_a \mid}^2+\lambda_{12} {\mid \phi_1 \mid}^2 {\mid \phi_2 \mid}^2+\tilde{\lambda}_{ab} {\mid \phi_a^\dagger \tilde{\phi}_b \mid}^2+\tilde{\lambda}_{0a}^{\prime}  \phi_a^{\dagger} H H^{\dagger}\phi_a \cr
&&+\lambda_{3} (\phi_1^\dagger H \phi_2^\dagger H +{\it{h.c.}})
\end{array}
\label{eq:hpotential}
\end{equation}
where $a=1,2$. The terms proportional to $\mu_{0a}$ and $\mu_{12}$ are $\mathbb{Z}_5$ soft breaking terms that provide the correct electromagnetic invariant vacuum, whereas $\eta$ takes care of  the necessary Lepton number violation in order to generate Majorana neutrino masses (radiatively). Due to its heavy mass and small mixing, $\eta$ does not play a relevant  role in the phenomenology of the Higgs potential.  Besides, as it was  shown in \cite{Aranda:2012bv},  the model can satisfy the strong flavor changing neutral current constraints coming from $K-\overline{K}$ mixing.

In order to obtain a stable vacuum (bounded from below) the following conditions are needed:
\begin{eqnarray}
\lambda_0, \lambda_a> 0 , \,\, \,\, \lambda_{12} +\tilde{\lambda}_{12} +\tilde{\lambda}_{21} > 2 \sqrt{\lambda_1 \lambda_2}, \,\, \,\, \lambda_{0a}+\lambda'_{0a} > - 2 \sqrt{\lambda_a \lambda_0}, \,\, \,\, a=1,2
\end{eqnarray} 
which are taken into account in the numerical calculation of the Higgs spectrum.

The  Higgs doublet fields are expressed as:
\begin{eqnarray}
H=\left(\begin{array}{c}\phi_0^+ \\ \displaystyle{\frac{v_0+r_0+iz_0}{\sqrt{2}}}\end{array}\right), \hspace{7mm} \Phi_a=\left(\begin{array}{c}\phi_a^+ \\ \displaystyle{\frac{v_a+r_a+iz_a}{\sqrt{2}}}\end{array}\right) .
\label{eq:hfields}
\end{eqnarray}
For simplicity we assume that there is no CP violation in the Higgs sector. 

\subsection{Higgs mass matrices}

The CP-even Higgs mass-squared matrix can be written as
\begin{equation}
 S^{2}_{ij}=\begin{pmatrix}
\displaystyle{-\frac{2 \mu_{12}^2 v_2+2 \mu_{01}^2 v_0+\lambda_3 v_2 v_0^2}{2 v_1}}& 
\mu_{12}^2-\frac{1}{2}\lambda_3 v_0^2 &\mu_{01}^2+ \lambda_3 v_2 v_0 \cr
\displaystyle{\mu_{12}^2-\frac{1}{2} \lambda_3 v_0^2}& 
-\frac{2 \mu_{12}^2 v_1+2 \mu_{02}^2 v_0+\lambda_3 v_1 v_0^2}{2 v_2}&
\mu_{02}^2+ \lambda_3 v_1 v_0 \cr
\mu_{01}^2+ \lambda_3 v_2 v_0 &\mu_{02}^2+ \lambda_3 v_1 v_0&\displaystyle{-\frac{ \mu_{01}^2 v_1+v_2(\mu_{02}^2 +2\lambda_3 v_1 v_0)}{v_0}} 
\end{pmatrix} ,
\label{eq:phmatrix}
\end{equation}
and the physical states are $(H_0, \,H_1,\, H_2)$ given by $H_i = U^{even}_{ij} {\cal{H}}_j$, where $H_0 =h$ is the SM-like Higgs.

Similarly, the charged  Higgs matrix  is given by
\begin{eqnarray}
C^2_{ij} =  S^2_{ij} + \Delta C^2_{ij} ,
\end{eqnarray}
where
\begin{equation}
 \Delta C^2_{ij} =\frac{1}{2}\begin{pmatrix}
\displaystyle{v_2^2 (\tilde{\lambda}_{12}+\tilde{\lambda}_{21})-v_0^2 \lambda'_{01} } & 
  v_0^2 \lambda_{3} - v_1 v_2 (\tilde{\lambda}_{12}+\tilde{\lambda}_{21})&
 v_1 v_0 \lambda'_{01}  - v_2 v_0 \lambda_{3} \cr
 v_0^2 \lambda_{3} - v_1 v_2 (\tilde{\lambda}_{12}+\tilde{\lambda}_{21})&
\displaystyle{v_1^2 (\tilde{\lambda}_{12}+\tilde{\lambda}_{21})-v_0^2 \lambda'_{02} } 
&v_2 v_0 \lambda'_{02}  - v_1 v_0 \lambda_{3} \cr
 v_1 v_0 \lambda'_{01}  - v_2 v_0 \lambda_{3} &v_2 v_0 \lambda'_{02}  - v_1 v_0 \lambda_{3} &\displaystyle{2 v_2 v_1 \lambda_{3}  - v_1^2 \lambda'_{01}-v_2^2 \lambda'_{02}} 
\end{pmatrix}.
\label{eq:shmatrix}
\end{equation}
From this mass matrix one gets the four physical states $(H_{1}^\pm, \, H_{2}^\pm)$ and the two charged Goldstone bosons $ G_W^\pm$,
where $H^\pm_i = V_{ij} \phi^\pm_j$ 
$i, \, j = 0,\,  1, \, 2$, and the matrix elements $V_{ij}$ are calculated numerically.

The CP-odd mass matrix then becomes
\begin{eqnarray}
R^2_{ij} =  S^2_{ij} + \Delta R^2_{ij} ,
\end{eqnarray}
where
\begin{equation}
 \Delta R^2_{ij}=\begin{pmatrix}
2  v_1^2 \lambda_{1}& 
\lambda_{3} v_0^2+v_1 v_2 \lambda_{12} &
v_1 v_0 ( \lambda_{01} + \lambda'_{01})  \cr
\lambda_{3} v_0^2+v_1 v_2 \lambda_{12}
&2  v_2^2 \lambda_2&
v_2 v_0 ( \lambda_{02} + \lambda'_{02}) \cr
v_1 v_0 ( \lambda_{01} + \lambda'_{01}) 
&v_2 v_0 ( \lambda_{02} + \lambda'_{02})  &
2 (\lambda_0 v_0^2 + v_1 v_2 \lambda_3) 
\end{pmatrix}.
\label{eq:chmatrix}
\end{equation}
From this mass matrix one obtains two pseudoscalar physical states ($A_1$, $A_2$) and one neutral Goldstone boson $A_0 = G_Z$, with
$A_i = U^{odd}_{ij} {\cal{H}}_j$ $i, \, j = 0,\,  1, \, 2$.

One can see that, if we were to decouple the scalar field $H$ from the other Higgs doublets, it would be possible to recover a 2HDM model.
%%%%%%%%%%%%%%%%
%\subsection{Numerical analysis of Higgs spectrum}
%%%%%%%%%%%%%%%%

\begin{figure}[h]
%\subfigure[]{\includegraphics[height=5cm]{espectro3d.pdf}}
%\subfigure[]{\includegraphics[height=5cm]{espectroplano.pdf}}
\includegraphics[height=7.6cm]{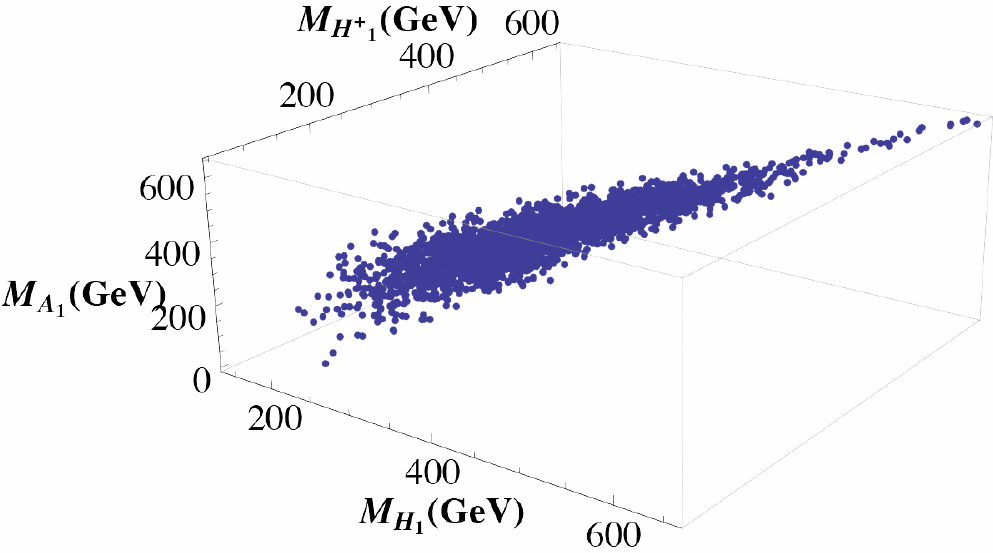}\\ \vspace*{0.5cm}
\includegraphics[height=8cm]{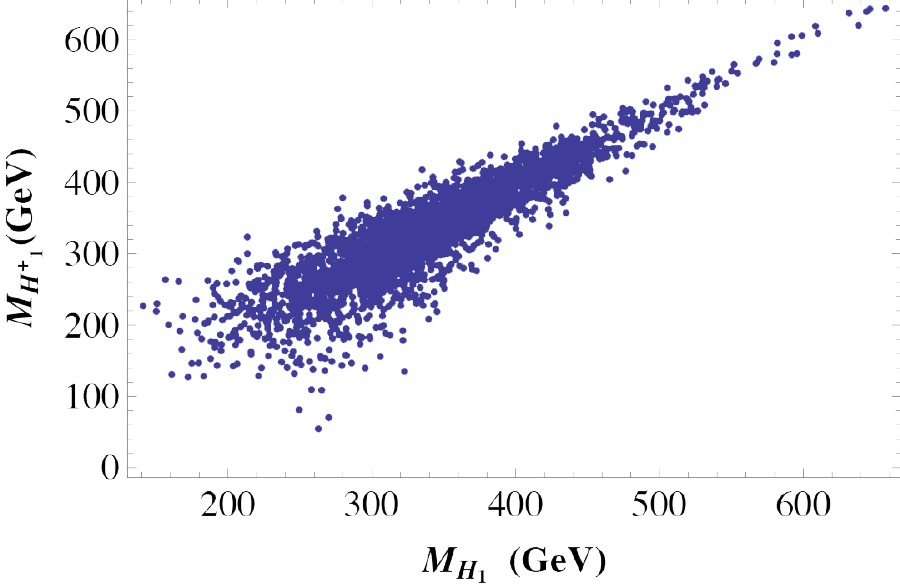}
\caption{ (Top) Higgs mass values $M_{H_1}\, \textrm{vs} \, M_{H^{+}_1}\, \textrm{and} \,M_{A_1}$ (in GeV).  (Bottom) Correlation between $M_{H_1}\, \textrm{and} \,M_{H^{+}_1}$ (in GeV) for $M_{A_1}=300$ GeV. The points correspond to values of the parameters that satisfy $K-\overline{K}$ mixing constraints and a light Higgs mass in the range of $124 - 126$~GeV.}
\label{fig:higgsmasses}
\end{figure}

A scan of the scalar potential parameter space was performed setting all dimensionless parameters $|\lambda|<1$ and fixing the mass of of the lightest CP even to be in the $124-126$~GeV range. Sets of parameters were singled out that comply with $K-\overline{K}$ mixing constraints. The plots in this paper correspond to one such set of parameter values. In Figure~\ref{fig:higgsmasses} we plot  $M_{H_1}\, \textrm{vs} \,M_{H^{+}_1}\, \textrm{and} \,M_{A_1}$ (top plot). We can see a wide region where the masses are in the (quasi decoupled) range given by $280-680$~GeV. More interesting are the points corresponding to lighter masses: taking a slice in the $M_{H_1} - M_{H^{+}_1}$ plane at $M_{A_1} = 300$~GeV (bottom plot), we find some such points  with $M_{H_1} < 200$~GeV and $M_{H_1^+} \sim 120-280$~GeV that satisfy all bounds and are consistent with all existing data.

%

%%%%%%%%%%%%%
\section{ \label{sec:yukawasector} Yukawa sector}
%%%%%%%%%%%%%

The Yukawa sector of this model is given by
\begin{eqnarray}
{\cal L}^{\bar{f}_i f_j \phi} & =&
-\left\{\frac{\sqrt2}{v}\overline{u}_i
\left(m_{d_j} X_{ij}^a {P}_R+m_{u_i} Y_{ij}^a {P}_L\right)d_j \,H^+_a
+\frac{\sqrt2m_{{l}_j} }{v} Z_{ij}^a\overline{\nu_i^{}}P_R{l}_j^{}H^+_a
+{H.c.}\right\} \nonumber \\
 & & -\frac{1}{v} \bigg\{  m_{f_i} h_{ij}^f (\bar{f}_iP_L f_j +\bar{f}_jP_R f_i )h + m_{f_i} H_{ij}^{af}   (\bar{f}_iP_L f_j +\bar{f}_jP_R f_i )H_a 
\\ \nonumber & &- i  m_{f_i} A_{ij}^{af}  (\bar{f}_iP_L f_j -\bar{f}_jP_R f_i )  A_a\bigg\},
\label{eq:lagrangian-f}
\end{eqnarray}
where $X_{ij}^a$, $Y_{ij}^a$  are the Yukawa quark couplings to $H_a^+$ for each chirality, $Z_{ij}^a$ are the Yukawa lepton couplings to $H_a^+ $ ($a=1,2$), and $f_i$ ($m_{f_i}$) is the $i$-th fermion field (mass). The generic Yukawa fermionic ($ffh$) couplings  are denoted by $h_{ij}^f$, whereas $H_{ij}^a$, $A_{ij}^{af}$ denote the corresponding couplings to $H_a$ and   $A_a$ respectively. The  interactions $f_i f_j  \phi$ induce  FCNC and include the information of  the Yukawa texture chosen in our model.  This generic interaction Lagrangian can be found in Ref.~\cite{HernandezSanchez:2012eg}.

\begin{figure}[h]
%\subfigure[]{\includegraphics[height=5cm]{acoplo_bottom.pdf}}
%\subfigure[]{\includegraphics[height=5cm]{acoplo_tau.pdf}}
\includegraphics[height=8cm]{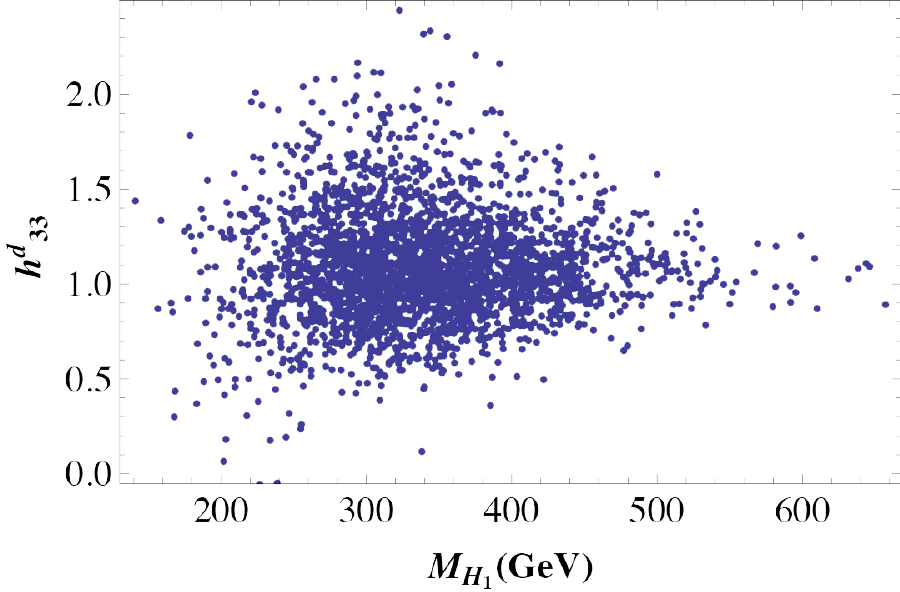} \\ \vspace*{0.5cm}
\includegraphics[height=8cm]{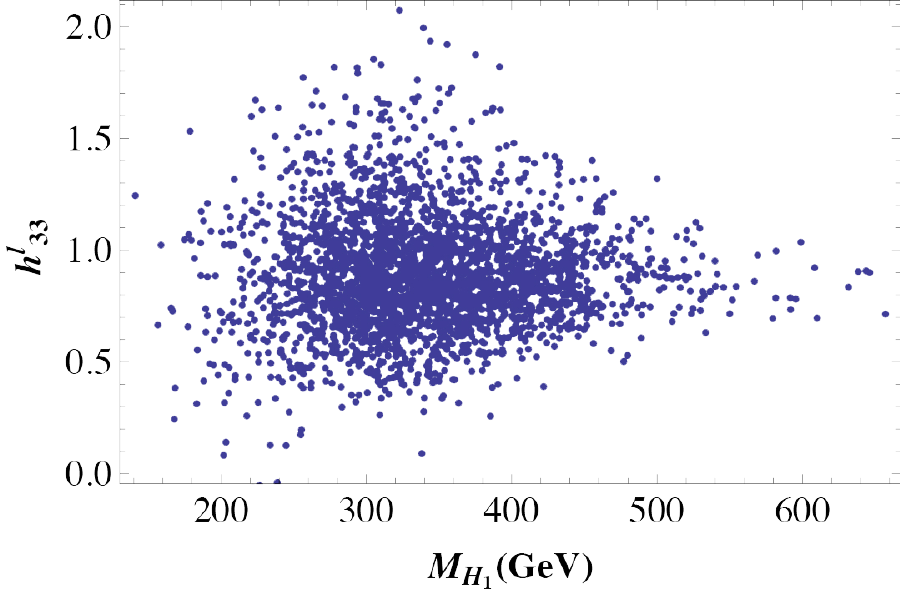}
\caption{(Top) $h_{33}^d \, \textrm{vs} \, M_{H_1}$ and (bottom) $h_{33}^l \, \textrm{vs} \,M_{H_1}$. The points correspond to values of the parameters that satisfy $K-\overline{K}$ mixing constraints and a light Higgs mass in the range of $124 - 126$~GeV.}
\label{fig:hm}
\end{figure}

Following the recent analysis  of the channel decay 
$h\to \gamma \gamma$ \cite{Aranda:2013kq}, we select two possibilities to obtain the bounds for the  decay rate of this mode: (a) the loop contribution of a light charged Higgs or (b) the reduction of the coupling $h$ to bottoms. In Figure \ref{fig:hm} we show $h_{33}^d \, \textrm{vs} \, M_{H_1}$ and $h_{33}^l \, \textrm{vs} \,M_{H_1}$. In the plot on top one can see that a hight  density around $M_{H_1} \sim 300 \textrm{GeV}$ with $h_{33}^d \sim 1$,  and at the bottom plot the same is shown for  $h_{33}^l \sim 0.8$. All the points of this region satisfy the light Higgs mass bound of  $\sim125 \, \textrm{GeV}$, therefore, there are several points that contain an allowed region with both scenarios (a) and (b). These points will be important when we study  flavor physics processes.
%

%%%%%%%%%
\section{ \label{sec:flavorconstraints} Flavor constraints on the 3HDM with flavor symmetry}
%%%%%%%%%

In this section we discuss the main flavor constraints for our model, considering scenarios where there is one SM-like Higgs boson $h^0$.

\subsection{$\mu - e$ universality in $\tau$ decays}

\begin{figure}[h!]
{\includegraphics[height=8cm]{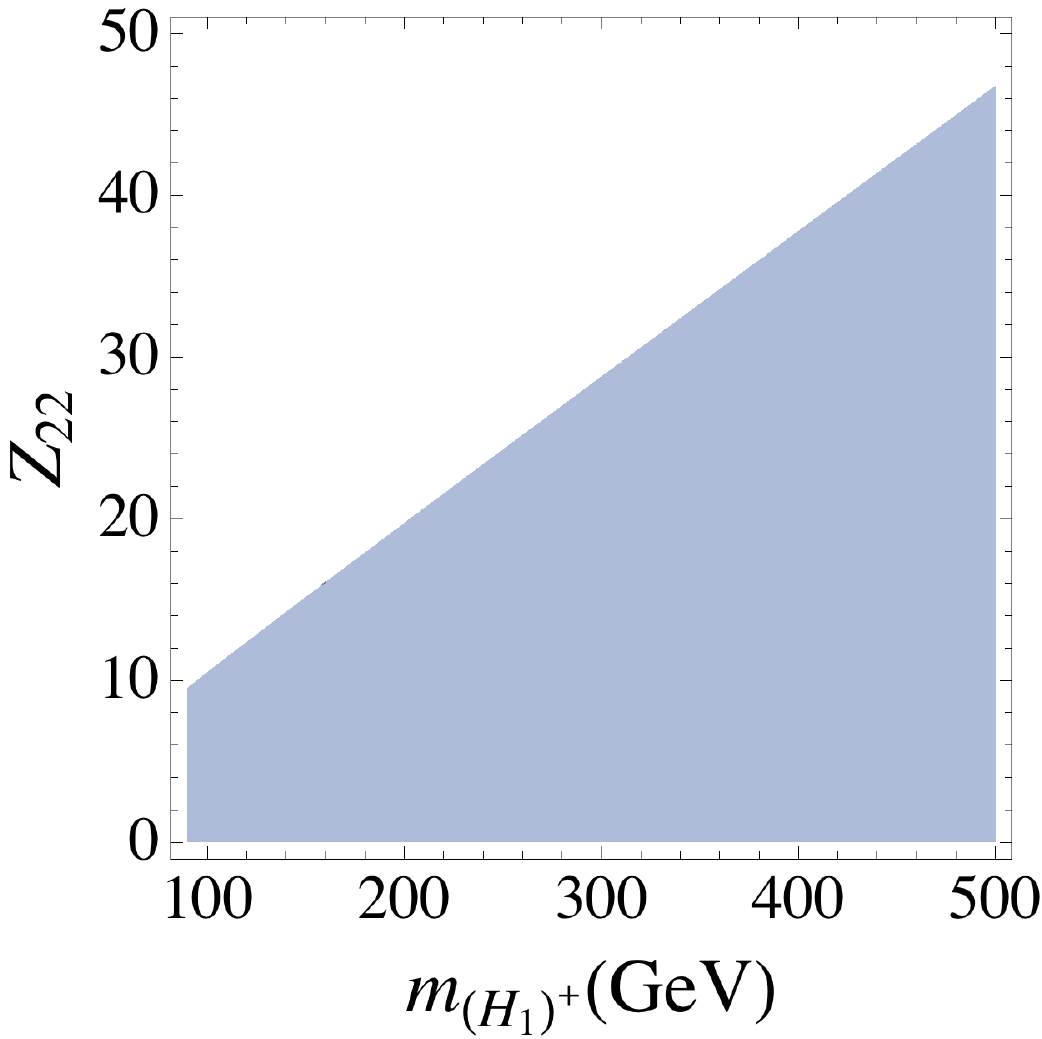}}
\caption{Allowed region  in the $m_{H_1^\pm} - Z_{22}^1$ plane, taking  $m_{H^\pm_1} \sim m_{H^\pm_2}$, $Z_{22}^1 \sim Z_{22}^2$ and $Z_{33} \sim 10 \times Z_{22}$, coming from the constraints of $\tau$ decays. }
\label{fig:deltaub1}
\end{figure}

The experimental results  for  $\tau \to \mu \bar{\nu}_\mu \bar{\nu}_\tau, \,\, e \bar{\nu}_e \bar{\nu}_\tau$ can be quantified by
\begin{eqnarray}
\bigg(\frac{g_\mu}{g_e}\bigg)^2 = \frac{BR(\tau \to \mu )  f(x_e)}{BR(\tau \to e) f(x_\mu)} = 1.0036 \pm 0.0020
\end{eqnarray}
where $x_\ell = m_\ell^2/m_\tau^2$ and $f(x)=1-8x^2+8x^3 -x^4 -12 x^2 \log{x}$.  This measurement  imposes constraints to the charged Higgses masses and couplings to leptons leading to the following bound~\cite{Jung:2010ik,HernandezSanchez:2012eg, Grossman:1994jb}:
\begin{eqnarray}
\sum_a \frac{Z_{22}^a Z_{33}^a }{m_{H_a^\pm}^2}  \leq 0.16  \,\,\,\, (95\% C.L.) 
\label{c1}
\end{eqnarray}
One can see that when $m_{H^\pm_1} << m_{H^\pm_2}$, the second charged Higgs is decoupled and we recover the result of  Ref \cite{ Grossman:1994jb}.
In our model, the numerical values of $Z_{ii}^a$ for the set shown in  figure \ref{fig:higgsmasses}, are $Z_{22}^a \sim 0.1$ and  $Z_{33}^a \sim O(1)$, in both cases the bound is avoided when $m_{H^\pm_1} \sim m_{H^\pm_2}$ and   $m_{H^\pm_1} < m_{H^\pm_2} $. Besides, we can obtain  a general allowed region from the constraint (\ref{c1}). This is shown in  Figure~\ref{fig:deltaub1}. We see that  the most constrained region is for the case 
$m_{H^\pm_1} \sim m_{H^\pm_2}$, $Z_{22}^1 \sim Z_{22}^2$  and $Z_{33}^1 \sim Z_{33}^2$, with $Z_{33} \sim 10 \times Z_{22}$.   Assuming this, one obtains the following bound: $|Z_{22}|< 20$ ($|Z_{22}|< 45$) for  $m_{H^\pm_1} < 250$ GeV ( $m_{H^\pm_1} < 500$ GeV). When the second charged Higgs is decoupled  the upper limit for $|Z_{22}| $  increases to 80. 
%%%%%%%%%
\subsection{Meson decays $M \to \ell \nu_\ell$}
%%%%%%%%%

As discussed in~\cite{Deschamps:2009rh,Jung:2010ik,HernandezSanchez:2012eg}, the total leptonic decay width of a charged meson,  due to the helicity suppression of the SM amplitude,  is sensitive to the $H^+$ exchange. The total decay width is given by
\begin{equation}
\label{eq:mlgamma}
\Gamma(M_{ij} \rightarrow l \nu)=\frac{G_F^2 m_l m_{M_{ij}}}{8 \pi}
 f_M^2 {\mid V_{ij} \mid}^2 (1+\delta_{em}){\mid 1- \Delta_{ij} \mid},
\end{equation}
where $i,j$ represent the valence quarks of the meson, $V_{ij}$ is the relevant CKM matrix element, $f_M$ is the decay constant of the meson $M$, $\delta_{em}$ denotes the electromagnetic radiative contributions, and $\Delta_{ij}$ is the correction that comes from new physics~\cite{Jung:2010ik,Deschamps:2009rh}
\begin{equation}
\Delta_{ij}=\sum_{n}\left( \frac{m_M}{m_{H^{\pm}_n}} \right)^2 Z_{kk} \left( \frac{Y_{ij}m_{u_i}+X_{ij}m_{d_j}}{V_{ij}(m_{u_i}+m_{d_j})} \right), \quad
k=2,3.
\label{eq:gendelta}
\end{equation}
This new physics contribution can be a complex number (like in 2HDM-III). In this work we are interested on the analysis of the function  $\Delta_{ij}$ for the measured heavy pseudoscalar mesons decays $B\to \tau \nu$,  $D\to \mu \nu$  and $D_s\to \mu \nu, \tau \nu$. We perform the analysis by using the expression for $\Delta_{ub,cd,cs}$ that corresponds to each process, and neglecting the contribution from any light quark mass ($m_u/m_b\leq m_d/m_c \sim$ O($10^{-3}$) ): 
\begin{equation}
\Delta_{ub}\simeq \sum_{n=1}^2 \left( \frac{m_B}{m_{H_n^{\pm}}} \right)^2 Z_{22}^n \left( \frac{X_{13}^n}{V_{ub}} \right),
\label{eq:deltaub}
\end{equation}
\begin{equation}
\Delta_{cd} \simeq  \sum_{n=1} ^2\left( \frac{m_D}{m_{H_n^{\pm}}} \right)^2 Z_{22}^n \left( \frac{Y_{21}^n}{V_{cd}} \right),
\label{eq:deltacd}
\end{equation}
\begin{equation}
\Delta_{cs}\simeq \sum_{n=1}^2 \left( \frac{m_D}{m_{H_n^{\pm}}} \right)^2 Z_{kk} \left( \frac{Y_{22}^n m_c + X_{22}^n m_s}{V_{cs} (m_c+ m_s)} \right), \quad
k=2,3.
\label{eq:deltacs}
\end{equation}
All numerical values ($V_{CKM}$ entries, quark and meson masses) are taken from~\cite{Beringer:1900zz}. 
We consider two relevant scenarios:  (I) $m_{H_1^{\pm}}\sim m_{H_2^{\pm}} $ and (II) $m_{H_1^{\pm}}<m_{H_2^{\pm}} $, in particular $m_{H_1^{\pm}}=150 \, \textrm{GeV}$ and $m_{H_2^{\pm}}=300 \, \textrm{GeV}$. When both charged Higgses are heavy, their contribution decouple and they can be neglected. As shown in our previous work~\cite{Aranda:2013kq}, the coupling $ff \phi$ could be modified as a consequence of the Yukawa texture,  and the combination of this exotic physics with the effects of the Higgs potential,  can relax the constraints for the parameter space obtained in other models with three Higgs doublets (see e.g. the "democratic 3HDM"~\cite{Grossman:1994jb}). In this work we are interested, as mentioned above, on the possibility of having light Higgses.

\begin{figure}[h!]
{\includegraphics[height=8cm]{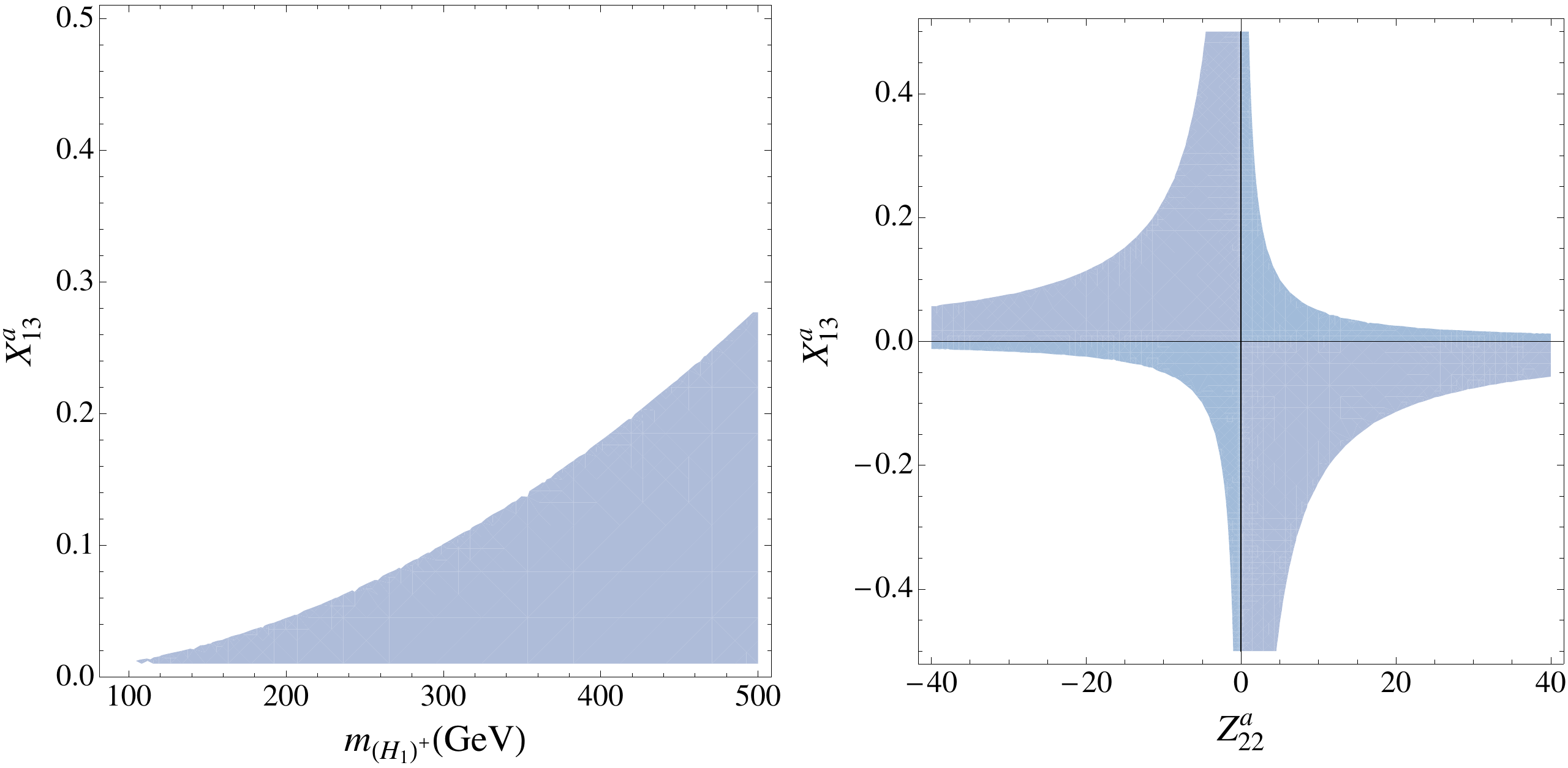}}
\caption{ Allowed region in the $ m_{H_1^{\pm}} - X_{13}^a$~(left) and $Z_{22}^a - X_{13}^a$~(right)  planes, corresponding to $B \to \tau \nu$, taking  $m_{H^\pm_1} \leq m_{H^\pm_2}$ and $Z_{22}^1 \sim Z_{22}^2$.}
\label{fig:deltaub}
\end{figure}

\begin{figure}[h!]
{\includegraphics[height=8cm]{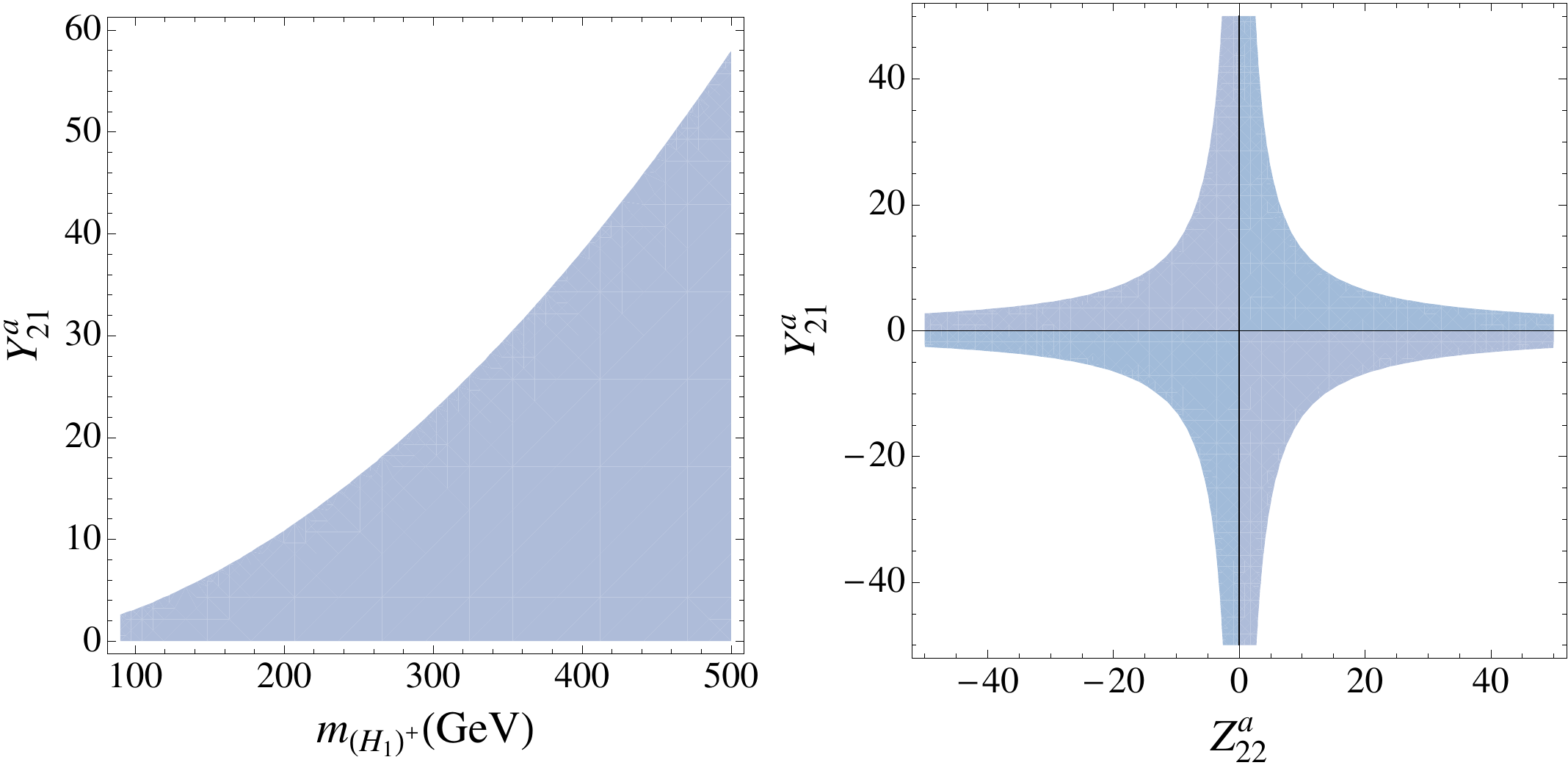}}
\caption{ Allowed region in the $ m_{H_1^{\pm}} - X_{13}^a$~(left) and $Z_{22}^a - X_{13}^a$~(right) planes, corresponding to $D \to \mu \nu$ and taking  $m_{H^\pm_1} \leq m_{H^\pm_2}$ and $Z_{22}^1 \sim Z_{22}^2$.}
\label{fig:deltacd}
\end{figure}

%%%%%%%%%%%%%%%%%%%
%\subsubsection{Numerical analysis of $\Delta_{ij}$}
%%%%%%%%%%%%%%%%%%%
In order to obtain constraints from the parameters $\Delta_{ij}$,  we consider the most recent experimental results. First we take the measurement from the Belle Collaboration of  a 4$\sigma$ signal  of  the branching fraction $B(B^- \to \tau^- \bar{\nu}) = 0.96 \pm 0.026 \times 10^{-4}$~\cite{Adachi:2012mm}, which is consistent with SM expectation. From this result we can get constraints on the parameters in equation~\eqref{eq:deltaub}:
$-$1.003  $\leq \Delta_{ub} \leq 0.222$  or  1.783 $\leq \Delta_{ub}  \leq 3.009$. On the other hand, considering the experimental results  of 
$B(D \to \mu \nu)$ given by the CLEO collaboration~\cite{Eisenstein:2008aa}, and the most recent measurement by the BESIII collaboration~\cite{Ablikim:2013uvu}, one can extract the following constraints: 
$-$0.129  $\leq \Delta_{cd} \leq 0.122$  or  1.868 $\leq \Delta_{cd}  \leq 2.127$. In the same way, using the experimental data of the CLEO collaboration for $B(D_s^- \to \ell^- \bar{\nu})$~\cite{Artuso:2007zg}, the constraint for 
$D_s \to \mu \nu, \, \, \tau \nu$ is  $-$0.170  $\leq \Delta_{cs} \leq 0.030$  or  1.903 $\leq \Delta_{cs}  \leq 2.088$. Using these experimental data and combining the results from $\mu - e$ universality in $\tau$ decays, we can get the most constrained allowed region for the $B\to \tau \nu$ ,  $D\to \mu \nu$  and $D_s\to \mu \nu, \tau \nu$ processes for $m_{H^\pm_1} \leq m_{H^\pm_2}$ and $Z_{22}^1 \sim Z_{22}^2$. For the decay $B\to \tau \nu$ we show in  Figure~\ref{fig:deltaub} the permitted region in both the $ m_{H_1^{\pm}} - X_{13}^a$  and $Z_{22}^a -X_{13}^a$ planes. One can obtain from the left panel the constraint $X_{13}^a <0.05$  ($X_{13}^a <0.3$)  for  $ m_{H_1^{\pm}} < 200$ GeV ($ m_{H_1^{\pm}} < 500$ GeV ). Note that the points shown in the Figure~\ref{fig:higgsmasses} correspond to $X_{13}^a \sim 0.01$ and so all of them are allowed  by $B\to \tau \nu$.   
Now, we carry out the same analysis  for the $B(D^+ \to \mu \nu)$ with the constraints  from $  \Delta_{cd} $. In this case the allowed parameter space is shown in Figure~\ref{fig:deltacd}, where we show it in the $ m_{H_1^{\pm}} - Y_{21}$  and $ Z_{22}^a - Y_{21}$ planes. The constraint becomes $Y_{21}^a <10$  ($Y_{21}^a <60$)  for  $ m_{H_1^{\pm}} < 200$ GeV ($ m_{H_1^{\pm}} < 500$ GeV ). Again, form the Higgs spectrum given in Figure~\ref{fig:higgsmasses}, we find that almost all points satisfy $Y_{21}^a <10$, and just a few points with $Y_{21}^a \sim 20$  are disfavored by the measurement of $D^+ \to \mu \nu$ for $m_{H_1^{\pm}} \leq 250$ GeV.

\begin{figure}[h!]
{\includegraphics[height=8cm]{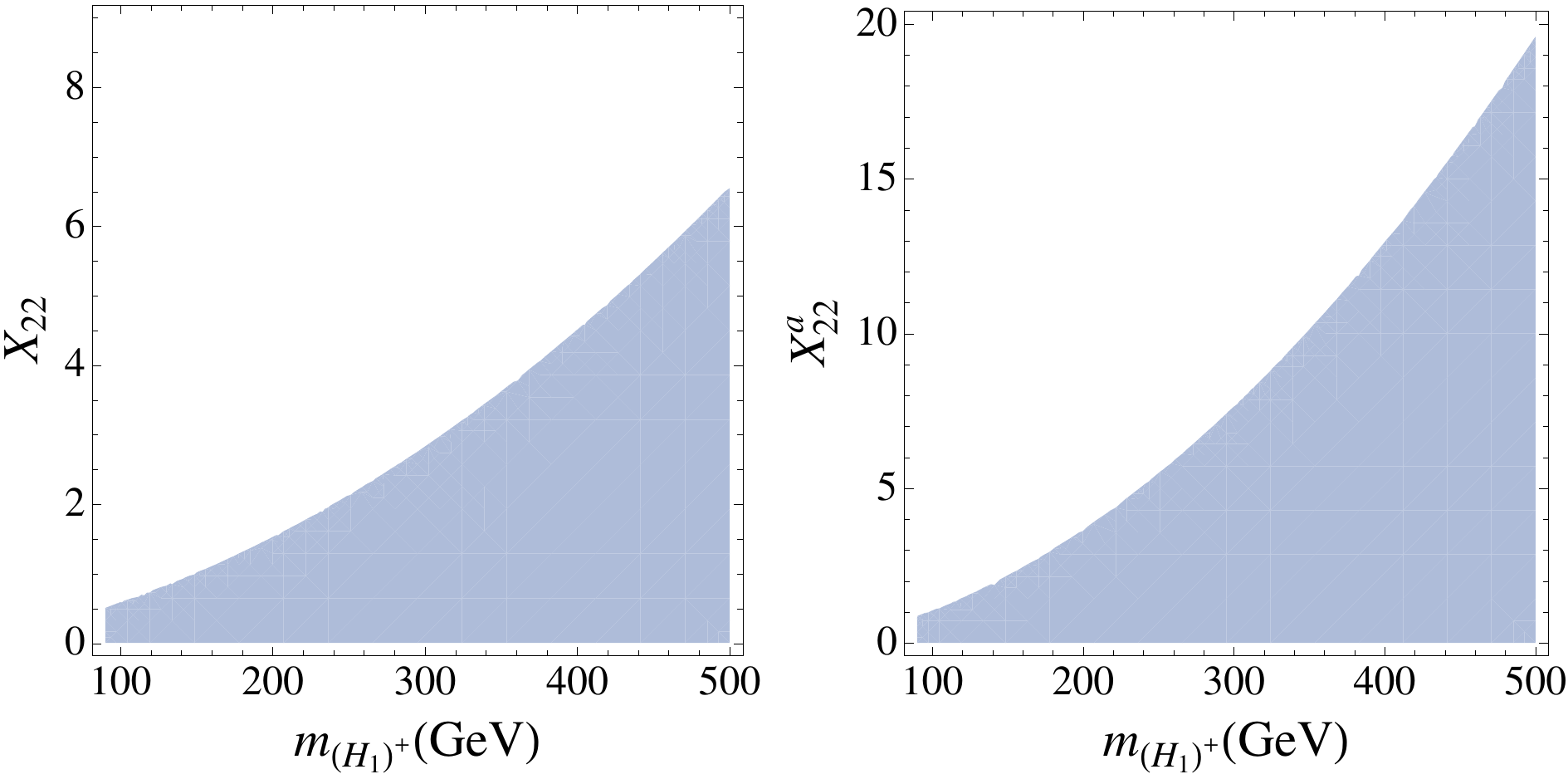}}
\caption{ Allowed region in the plane  $ m_{H_1^{\pm}} - X_{22}^1$~(left)  and $ m_{H_1^{\pm}} - X_{22}^a$~(right) planes corresponding to $D_s \to \ell \nu$, with $X_{22}^1 \sim Y_{22}^1 \leq  X_{22}^2 \sim Y_{22}^2 $~(left) and $Y_{22}^1 \sim Y_{22}^2 \leq  X_{22}^1 \sim X_{22}^2 $~(right). The plots are presented for $m_{H^\pm_1} \leq m_{H^\pm_2}$ and $Z_{22}^1 \sim Z_{22}^2 \leq 40$.}
\label{fig:deltacs}
\end{figure}

In Figure~\ref{fig:deltacs} we present the region allowed by the measurement $D_s\to \mu \nu, \tau \nu$ in the $ m_{H_1^{\pm}} - X_{22}^1$(leftl)  and $ m_{H_1^{\pm}}- X_{22}^a$ (right) planes taking $m_{H^\pm_1} \leq m_{H^\pm_2}$ and $Z_{22}^1 \sim Z_{22}^2 \leq 40$. According to the behaviour of the points in Figure~\ref{fig:higgsmasses}, we choose the cases with $X_{22}^1 \sim Y_{22}^1 \leq  X_{22}^2 \sim Y_{22}^2 $~(left) and $Y_{22}^1 \sim Y_{22}^2 \leq  X_{22}^1 \sim X_{22}^2 $ ~(right).  For the case on the left, one gets the bound $X_{22}^1\sim Y_{22}^1 \leq 2$ for $m_{H_1^\pm} \leq 250$ GeV ($X_{22}^1\sim Y_{22}^1 \leq 10$ for $m_{H_1^\pm} \leq 500$ GeV) and for the case on the right we obtain the following constraint: $X_{22}^a \leq 5$ for $m_{H_1^\pm} \leq 250$ GeV ($X_{22}^a \leq 10$ for $m_{H^\pm} \leq 500$ GeV) with $a=1,2$. 

\begin{figure}[h!]
{\includegraphics[height=8cm]{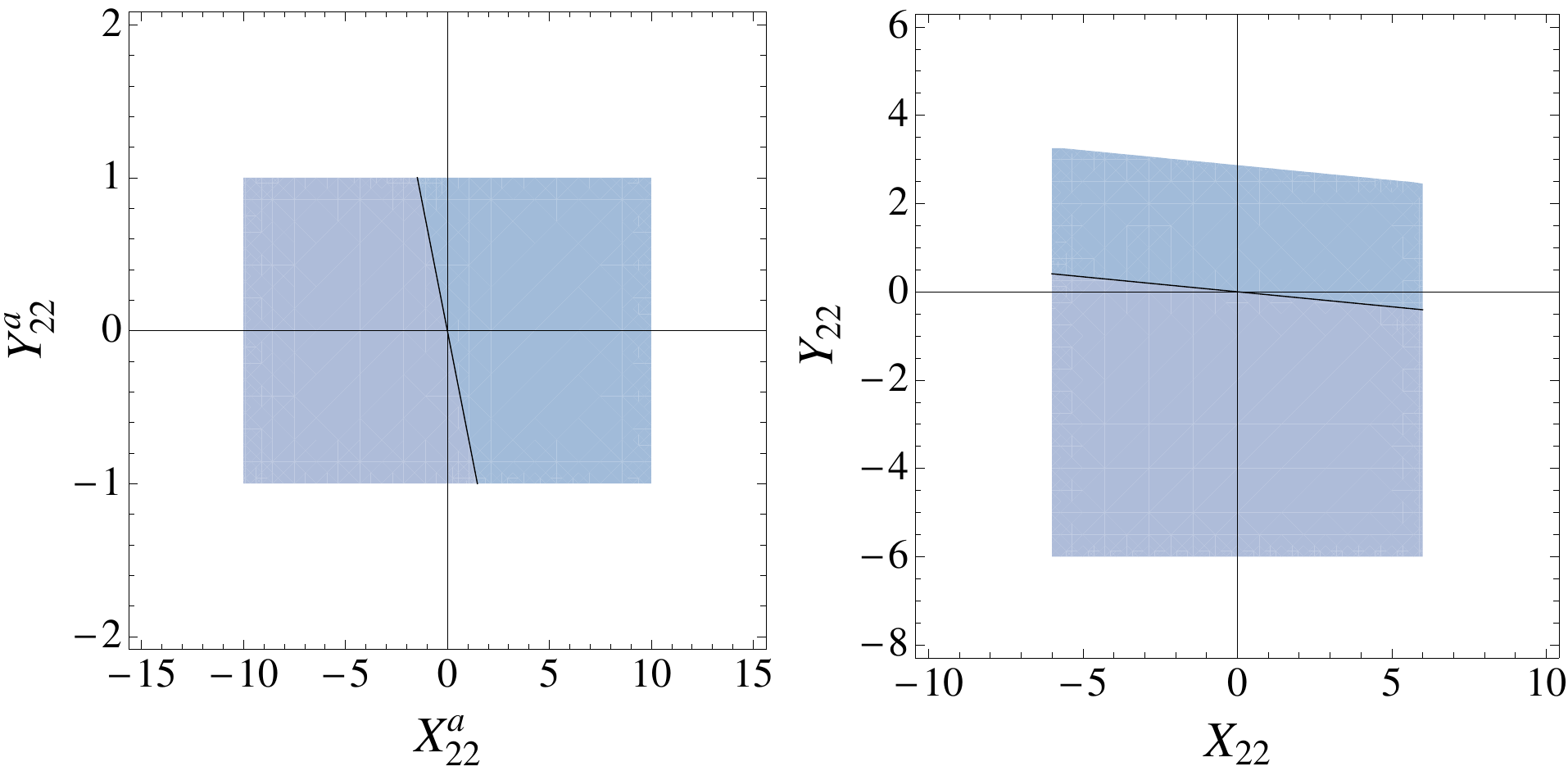}}
\caption{ Allowed region in the $X_{22}^a - Y_{22}^a$ plane for $Y_{22}^1 \sim Y_{22}^2 \leq  X_{22}^1 \sim X_{22}^2 $~(left) and in the $X_{22}^1 - Y_{22}^1$ plane for $X_{22}^1 \sim Y_{22}^1 \leq  X_{22}^2 \sim Y_{22}^2 $~(right), corresponding to $D_s \to \ell \nu$. The plots are presented for $m_{H^\pm_1} \sim 150 $ GeV and $m_{H^\pm_2}\sim 300$ GeV and $Z_{22}^1 \sim Z_{22}^2\sim 20$.}
\label{fig:deltacsb}
\end{figure}

In Figure~\ref{fig:deltacsb}, considering a combination of the results shown in Figure~\ref{fig:deltacs} with $m_{H_1^\pm} \leq 500$ GeV, we present a correlation among $X_{22}^a$ and $Y_{22}^a$ for the case $Y_{22}^1 \sim Y_{22}^2 \leq  X_{22}^1 \sim X_{22}^2 $ ~(left), where we obtain the bound $|Y_{22}^a|\leq 1$. For the case $X_{22}^1 \sim Y_{22}^1 \leq  X_{22}^2 \sim Y_{22}^2 $~(right) we present the $X_{11}^1 - Y_{22}^1$ plane and get the constraint $-6 \leq Y_{22}^1 \leq 3$. Note also that all points in Figure~\ref{fig:higgsmasses} survive the constraints and thus only the process $D^+ \to \mu \nu$  can eliminate some of them.

%%%%%%%%%
\subsection{Semileptonic decays $B \to D \tau \nu$}
%%%%%%%%%
The BaBar and Belle experiments have measured the branhing $B(B \to D \tau \nu)$ for first time~\cite{Aubert:2009at, Matyja:2007kt}. In particular, the BaBar collaboration has published the following ratios~\cite{Lees:2012xj}: \footnote{With this measurement the 2HDM type II  is disfavored, because it cannot explain $R(D)$ and $R(D^*)$ simultaneously, and for $B \to \tau \nu$ a large fine tuning is needed.}
\begin{eqnarray}
R(D) = 0.44 \pm 0.058\pm 0.042 \\ \nonumber
R(D^*)= 0.332\pm 0.024 \pm 0.018 \\
\end{eqnarray}
where $ R(D)= BR(B \to D \tau \nu)/ BR(B \to D \ell \nu)$.  According to the analysis in~\cite{Deschamps:2009rh}, one can calculate
the observable $R_{B \to D \tau \nu} =BR(B \to D \tau \nu)/ BR(B \to D e \nu)$, which corresponds to a $b \to c $ transition and is given as a second order polynomial of the $H_a^\pm c \bar{b}$ coupling:
\begin{eqnarray}
R_{B \to D \tau \nu} = a_0 + a_1 (m_B^2 - m_D^2) \delta_{23} + a_2 (m_B^2 - m_D^2)^2 \delta_{23}^2\ ,
\end{eqnarray} 
where the polynomial coefficients $a_i$ are given in  reference~\cite{Deschamps:2009rh} and  the factor $\delta_{23}$ is determined by
\begin{eqnarray}
\delta_{23} = - \sum_a \frac{Z_{33}^a}{m_{H_a^\pm}^2} \bigg( \frac{Y_{23}^a m_c - X_{23}^a m_b }{m_c- m_b} \bigg) \ .
\end{eqnarray}
\begin{figure}[h!]
{\includegraphics[height=8cm]{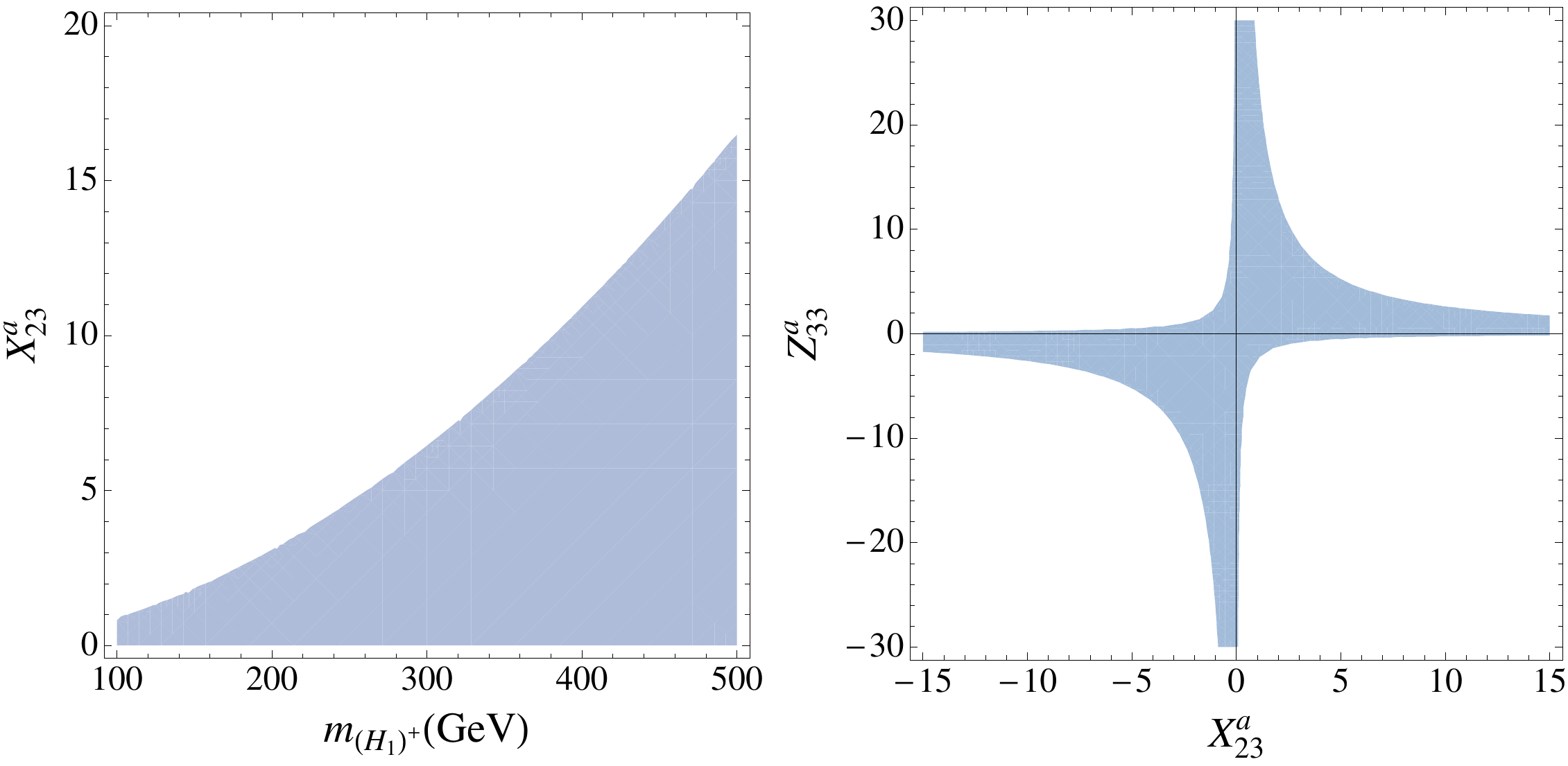}}
\caption{ Allowed region in the plane  $ m_{H_1^{\pm}}$ vs.  $X_{13}^a$  and $X_{23}^a$  vs.  $Z_{33}^a$, corresponding to $B \to D  \tau \nu$ decay, taking  $m_{H^\pm_1} \leq m_{H^\pm_2}$ and $Z_{33}^1 \sim Z_{33}^2$, with $X_{23}^a << Y_{23}^a$ .}
\label{fig:RD}
\end{figure}
In our study we consider both $ R(D)$ and $R(D^*)$ simultaneously. Figure~\ref{fig:RD} shows the allowed regions in the $ m_{H_1^{\pm}} - X_{23}^a$~(left)  and $X_{23}^a - Z_{33}^a$~(right) planes, with $m_{H^\pm_1} \leq m_{H^\pm_2}$ and $Z_{33}^1 \sim Z_{33}^2$, with $X_{23}^a << Y_{23}^a$. For these scenarios, we can derive  the bound $|X_{23}^a| \leq 4$ ($|X_{23}^a| \leq 16$ )  for $m_{H_1^\pm} \leq 250$ GeV  ($m_{H_1^\pm} \leq 500$ GeV). One can see, looking again at the points in  Figure~\ref{fig:higgsmasses},  that the 3HDM with a flavor symmetry can avoid the $ R_{B \to D \tau \nu}$ constraint with $X^a_{23}\sim 10^{-1}, 10^{-2}$ and $|Y^a_{23}| \sim 10$. This behavior of the couplings $X_{23}$ and $ Y_{23}$ can induce rather exotic physics, very different from 3HDM with NFC. In  particular, decays channels involving $H^\pm \to cb $ could be important in the transition $t \to H^\pm b$, when one charged Higgs $H^\pm$ is light enough~\cite{HernandezSanchez:2012eg, Akeroyd:2012yg, DiazCruz:2009ek}. 
%%%%%%%%%
\subsection{ $B \to X_s  \gamma$ decay }
%%%%%%%%%
According to the current average of the measurements by CLEO~\cite{Chen:2001fja}, BaBar~\cite{Aubert:2007my,Lees:2012wg,Lees:2012ym} and Belle experiments~\cite{Abe:2001hk,Limosani:2009qg},  
$BR(B \to X_s \gamma) _{E_\gamma} = 3.37 \pm 0.23 \times 10^{-4}$. Following the analysis for 3HDM presented in~\cite{ Chang:1991nr, Okada:1997qh,   Borzumati:1998tg, Kola:2006zv}, we can reproduce the LO contributions of both charged Higgses $H_a^{\pm}$ (with $a=1$ ,2) in our model. They are given by the relevant Wilson coefficients at the matching energy scale $\mu_W$:
\begin{eqnarray}
\delta C_{i}^{0,eff} (\mu_W) = \sum_a \bigg( \bigg|\frac{Y^a_{33} Y_{32}^{a*}}{V_{tb} V_{ts}^*} \bigg| C_{i, YY}^{0} (y_t^a)  +  
\bigg| \frac{X^a_{33} Y^{a*}_{32}}{ V_{tb} V_{ts}^*} \bigg|  C_{i, YY}^{0} (y_t^a) \bigg) \, \, (i=7, \, 8) \ ,
\end{eqnarray}
where $y_t^a = m_t^2/ m_{H^\pm_a}^2$ and the coefficients $C_i$ are very well known and can be found in~\cite{Borzumati:1998tg}. Moreover, the BR for the inclusive radiative decay $ B \to X_s  \gamma$ at the LO level is given by~\cite{ Chang:1991nr, Okada:1997qh,   Borzumati:1998tg, Kola:2006zv}:
\begin{eqnarray}
BR(B \to X_s  \gamma ) = B_{SL}  \frac{6 \alpha_{em}}{\pi \theta (z) \kappa (z)} \bigg| \frac{V_{tb} V_{ts}^*}{V_{cb}} \bigg|^2   
\bigg| C_{7}^{0,eff} (\mu) \bigg|^2 \ ,
\end{eqnarray}
where $B_{SL}= (10.74 \pm 0.16) \% $ of the semi-leptonic BR of the B meson, $\alpha_{em}$ is the fine-structure constant, $z = m_c^{pole}/ m_b^{pole}$ is the ratio of the quark pole masses, $\theta (z)$ and $\kappa (z)$ are the phase space factor and the QCD correction for the semileptonic B decay (given in~\cite{Borzumati:1998tg,Xiao:2003ya}). 
We assume that $|Y_{33}^a|$ is small (e. g. $|Y^a_{33}|<1$) as required by the low energy process  $Zb\bar{b}$~\cite{Degrassi:2010ne,Akeroyd:2012yg,HernandezSanchez:2012eg}. Then, following the  analysis of~\cite{Trott:2010iz,HernandezSanchez:2012eg}, and for  
80 GeV$\leq m_{H_a^\pm} \leq 360$ GeV, we obtain the constraints:
\begin{eqnarray}
\sum_{a=1}^2 \bigg|\frac{Y^a_{33} Y_{32}^{a*}}{V_{tb} V_{ts}^*} \bigg|  < 1, \,\,\,\,\,\,\,\,\,\, -1.7 < \sum_{a=1}^2 \bigg| \frac{X^a_{33} Y^{a*}_{32}}{ V_{tb} V_{ts}^*} \bigg| < 0.7 \ .
\end{eqnarray}
Practically all points in Figure~\ref{fig:higgsmasses} avoid these bounds.
%%%%%%%%%

\section{ \label{sec:conclusions} Conclusions}
In this work we have performed a thorough analysis of the 3HDM with a $\mathbb{Z}_5$ flavor symmetry paying special attention to regions of the parameter space where one can satisfy all flavor and collider constrains and at the same time having extra light states that could be discovered in the next run of the LHC.

We have identified regions with a charged Higgs $H^\pm$ around $150$ GeV where the contributions to different observables in the B-sector are under control. This leaves an interesting possibility to be tested at the LHC. It could be interested to further study the different signals of this light charged state.

\begin{acknowledgments}
This work was supported in part by CONACYT and PROMEP under the grant Red Tem\'atica: F\'isica del Higgs y del sabor. The work of AD was partially supported by the National Science Foundation under grant  PHY-1215979. JHS would like to thank support from VIEP-BUAP. 
\end{acknowledgments}

\bibliographystyle{apsrev4-1.bst}
\bibliography{Referencias.bib}

\end{document}